\documentstyle[12pt]{article}
\begin{document}

\begin{center} 

{\large 
Semiclassical Aspects of Quantum Mechanics \\

\vspace{0.2cm}
                       
                       by Classical Fluctuations} \\

\vspace{1.0cm}

Salvatore De Martino$^{*}$ \footnote{E--Mail: 
demartino@physics.unisa.it},
Silvio De Siena$^{*}$ \footnote{E--Mail: 
desiena@physics.unisa.it}, and
Fabrizio Illuminati$^{*}$ $^{a)}$ \footnote{E--Mail: 
fabrizio@spock.physik.uni-konstanz.de,
illuminati@physics.unisa.it} \\

\vspace{0.2cm}

* {\it 
Dipartimento di Fisica, Universit\`a di Salerno; \\
INFM, Unit\`a di Salerno; and INFN, Sezione \\
di Napoli -- Gruppo Collegato di Salerno, \\
I--84081 Baronissi (Salerno), Italy} \\

\vspace{0.3cm}

$a)$ {\it Fakult\"at f\"ur Physik, Universit\"at Konstanz, \\
D--78457 Konstanz, Germany}

\end{center} 

\vspace{0.3cm}

\begin{center} 21 January 1998 \end{center}

\vspace{0.3cm}

\begin{abstract} 
Building on a model recently proposed by F. Calogero,
we postulate the existence of a universal Keplerian tremor
for any stable classical system.
Deriving the characteristic unit of action 
$\alpha$ for each classical interaction, we obtain
in all cases $\alpha \cong h$,
the Planck action constant, suggesting that
quantum corrections to classical
dynamics can be simulated through a fluctuative
hypothesis of purely classical origin.
\end{abstract}

\vspace{0.2cm}


\vspace{0.4cm}

In a recent paper F. Calogero 
has put forward an intriguing conjecture on the possible
gravitational origin of quantization \cite{calogero}.

The scheme followed by Calogero is simple but appealing.
He suggests that the origin of quantization be attributed
to the universal interaction of every particle with the
background gravitational force due to all other particles
in the Universe. Such background interaction 
generates a chaotic component in the motion 
of each single particle, with
a characteristic constant $\tau$ measuring the time
scale of stochasticity (Zitterbewegung).

Assuming a basic granularity of the Universe, made up of
nucleons (or hydrogen atoms) of mass $m$, Calogero derives
an expression for Planck's constant, $h \cong G^{1/2}m^{3/2}R^{1/2}$, 
with $G$ the Newtonian
gravitational constant and 
$R$ the observed Radius of the Universe.
This formula, which connects the fundamental
constant of quantum theory with the fundamental
gravitational constants, was already known since
some time \cite{weinberg} but its meaning and
implications were so far unexplained; Calogero
provided the first derivation of it from a 
mechanical model, so that one could very well
name it the Calogero--Weinberg formula.

Now, the crucial point in the procedure carried out
by Calogero is that the characteristic
time $\tau$ of the stochastic 
motion per particle, being associated to 
a collective chaotic effect, should
be inversely proportional to the square 
root of $N$, the total number
of particles in the Universe \cite{calogero}: 
\begin{equation}
\tau \cong N^{-1/2}{\cal{T}} \, ,
\end{equation}

\noindent with ${\cal{T}}$ the characteristic global 
time unit associated with a
Universe of total mass $M$.
Defining the energy
per particle $\epsilon \cong E/N$, 
with $E$ total energy of the Universe,
and a global unit of action for the Universe 
$A=E{\cal{T}}$, Calogero defines
the unit of action per particle
\begin{equation}
\alpha = \epsilon \tau \cong N^{-3/2}A \, . 
\end{equation}

\noindent Replacing $N$ with 
the ratio of the global and
the granular amount of sources $M/m$, 
imposing that $\alpha$
be independent of extensive 
quantities and performing an elementary
dimensional analysis for the 
combination of the nucleon mass $m$,
the Radius of the Universe $R$ 
and the Newtonian gravitational
constant $G$, Calogero finally 
arrives at the expression 
\begin{equation}
\alpha \cong G^{1/2}m^{3/2}R^{1/2} \, ;
\end{equation}

\noindent inserting the numerical 
values $m \cong 10^{-27} kg$,
$G \cong 10^{-11} 
kg^{-1} \cdot m^{3} \cdot s^{-2}$
and the most updated cosmological
estimate for the observed Radius
of the Universe $R \cong 10^{30}m$
\cite{peebles}, eq.(3) yields 
$\alpha \cong h$, the Planck
action constant (we warn the reader
that in the presente work we are
neglecting in the numerical
computations all those factors
that do not substantially
affect the order of magnitude 
of the estimated quantities).

In conclusion, Calogero suggests that
quantization might be explained
via classical gravitation by assuming the existence of 
a chaotic component of the individual
particles' motion due to a ``universal
coherent tremor'' associated to the extremely
large number $N$ of the elementary components
making up the Universe. This conjecture is
mathematically implemented introducing 
Maxwell--Boltzmann fluctuations proportional to
$1/\sqrt{N}$.

In the present letter we take a closer look at
the model provided by Calogero. We first show that his
procedure can be equivalently reformulated by
replacing the fluctuative law eq.(1) with a
fractal space--time relation $l \sim \tau^{2/3}$
which is the mathematical expression of
a ``universal Keplerian tremor''.

We then move on to show that the scheme of Calogero
can be applied to all the other known interactions
(electromagnetic, strong, etc.) and that it leads
in all cases to a variety of formulas again linking 
Planck's constant with the proper fundamental constants
associated to each considered interaction.

In this way we achieve two purposes: on the one hand
we clarify that the fluctuative hypothesis of Calogero
actually holds also for systems with few degrees of 
freedom, since in our Keplerian
reformulation it does not involve the number $N$
of elementary constituents. On the other hand,
we show that the mechanism is universal, in the sense
that it allows to derive an expression for Planck's
constant for any physical system confined on the
typical space--time region associated to any of 
the fundamental interactions known in Nature.

In particular, one can apply the Keplerian
formulation of the Calogero model to ``Gedanken
Universes'' made of an arbitrary number $N$ of
gravitationally interacting particles, even 
with $N=2$. In any instance, one always
obtains $\hbar$ as the unit of action and the
observed Radius of our actual Universe as 
the typical length scale of the system,
irrespective of the assumed number $N$ of
elementary constituents.
Therefore the Keplerian fluctuation and its
relationship with the elementary quantum of
action $\hbar$ seem to have little or nothing
at all to do with a chaotic dynamics induced
by the enormous number $N$ of elementary 
constituents that interact gravitationally
in the actually observed Universe.

We can then draw the following conclusions:

1) The mechanism envisaged by Calogero seems to
capture some essential aspects of the interplay
between classical and quantum mechanics.

2) The universality of the mechanism as shown
in the present letter and its insensitivity to
$N$ being large or small strongly suggest that
the simple and appealing interpretation of
Calogero is untenable: quantum mechanics is
in fact the most fundamental theory known
up to now, and the derivation
of Planck's constant for each known interaction
by this procedure rules out the possibility
of a privileged role of classical gravitation as the
``origin'' of quantum mechanics.

3) Still it is remarkable that
a simple qualitative argument 
(the universal Keplerian tremor) allows to
capture some relevant quantum features by purely
classical considerations.

4) From the point of view of quantum mechanics 
we can then interpret Calogero's and our results
as a first embryonal, still qualitative, step towards
the possibility of recovering  semiclassical aspects
of quantum mechanics starting directly from classical
mechanics and implementing (simulating) quantum corrections
in terms of suitable, purely classical,
stochastic fluctuations.
This is at variance with the usual approximation
schemes of quantum mechanics, such as the WKBJ procedure,
which recover the semiclassical and classical domains
starting from the ``deep'' quantum domain. It is
rather a first operative definition of a semi--quantal
approximation scheme, hopefully 
to be further developed.

We begin by showing that the tremor 
hypothesis of Calogero, eq. (1) is equivalent 
to a generalization on the microscopic 
scale of Kepler third
law in the form $l \sim \tau^{2/3}$, where, by  
introducing the total volume of the 
Universe $V$ and the
mean allowed volume 
per particle (specific volume)
$v_{s}\cong V/N$, we have 
defined the mean free path 
of the individual constituents 
$l \cong v_{s}^{1/3}$.
In fact, we can immediately 
rewrite Calogero's fundamental
relation eq. (1) as:
\begin{equation}
\frac{{\cal{T}}^{2}}{V} \equiv
const. \cong \frac{\tau^{2}}{v_{s}} 
\sim \frac{\tau^{2}}{l^{3}} \, ,
\end{equation}

\noindent or, equivalently,
\begin{equation}
l \sim \tau^{2/3} \, .
\end{equation}

\noindent Note that, from $V \sim R^{3}$, the 
first member of eq. (4) is Kepler third 
law on the scale of the Universe.

One should note that
the most updated cosmological
scenarios \cite{peebles} lead to a 
recessing away law of galaxies
in the expanding Universe of the form
$L \sim t^{2/3}$ (with $L$ the
distance between galaxies). 
The congruence of this phenomenon
on large cosmological scales 
with our Keplerian version eq. (5)
of Calogero's tremor 
hypothesis eq. (1) implies the
extension of validity 
of a Kepler--like third law for
gravitational interactions ranging 
from small to large scales.
The resulting picture clearly
ignores the structure of the system in its
finest details, being  based on a sort of
mean field description. 

At this point we might draw the 
first provisional conclusion: if one
assumes, as usual, that
quantum mechanics is the fundamental
theory, then the above analysis implies
that the stability of the Universe
on the scale of its observed Radius $R$
is ruled, via the Calogero mechanism
as formulated in eq. (5),
by the Planck quantum of
action, in complete analogy with the 
stable confined systems associated
to the other known interactions.

This interpretation is of course totally
at variance with the one originally
given by Calogero, and to see whether it
is tenable, we should move on to apply
his model to the other stable systems
on different scales that are associated
to the other relevant known
interactions beyond gravity.

We expect in this way to obtain again formulas
linking Planck's action constant to the typical
radius $R$ of stability 
and to the fundamental masses and
interaction constants associated to
the systems being considered.

As we will show below, it turns out that this
is indeed the case. This fact seems to imply
that a simple qualitative picture based on
purely classical tools allows to recover
some fundamental aspects of quantum mechanics
in its semiclassical domain
in connection with the stability of matter.

We thus proceed to apply the scheme already adopted for
gravitation to aggregates
of charged particles interacting electromagnetically
and to systems of confined quarks in the
nucleons. We do this by assuming 
the tremor hypothesis in the
form of eq.(5) to hold for any stable
aggregate of particles, 
relying on the fact that for these
confined systems one can certainly introduce well
defined characteristic global units of time 
${\cal{T}}$ and volume $V$.

\vspace{0.2cm}

{\it Electromagnetic Interactions}.
Let us consider first the case of a stable
aggregate of charged particles 
interacting via electromagnetic forces.
The fundamental constants involved are
the electrostatic constant
$K \equiv 1/4\pi\epsilon_{0} \cong 10^{10} 
N\cdot m^{2}\cdot C^{-2}$, the elementary
unit of charge $e \cong 10^{-19}C$ and the
velocity of light $c \cong 10^{8} m \cdot 
s^{-1}$.

Since such aggregates are in 
general made of collections of 
electrons and protons, we can 
take as the natural unit of 
elementary mass the reduced mass $\mu$, 
which substantially coincides with the
mass of the electron $m \cong 10^{-30} kg$.
Let us consider also the characteristic linear 
dimension $R$ of the stable aggregate; 
this characteristic global scale of length 
can vary from $R \cong 10^{-2} m$ (macroscopic 
dimensions) to $R \cong 10^{-10} m$ (atomic 
dimensions).

Expressing $N$ as the ratio of the global and
the granular 
amount of sources $Q/e$, with $Q$ total charge of the
aggregate, imposing eq. (2) and 
requiring the independence of the unit of action
$\alpha$ on extensive quantities, we obtain 
$A=Q^{3/2}{\tilde A}$;
by dimensional considerations 
${\tilde A} = f(K,m,c,R)$, and we finally arrive at
\begin{equation}
\alpha \cong e^{3/2}K^{3/4}m^{1/4}c^{-1/2}
R^{1/4} \, .
\end{equation}

Inserting numbers in eq. (6) we have then in all cases,
up to at most one order of magnitude, 
$\alpha \cong 10^{-34} J \cdot s \cong h$, i.e.,
once more, Planck constant.

\vspace{0.2 cm}

{\it Quarks}.
We now move on to 
consider a hadron having as granular 
costituents a collection of bound quarks. 
The interaction we consider is
the ``string law" described by the typical confining 
potential $V = kr$  with the strength constant $k$
varying in the range $k \cong 0.1 GeV \cdot fm^{-1}
\div 10 GeV \cdot fm^{-1}$ (values compatible
with the experimental bounds \cite{perkins}).
Let us also introduce the quark masses 
$m \cong 0.01 GeV \cdot c^{-2}
\div 10 GeV \cdot 
c^{-2}$ \cite{particledata}, 
the velocity of light $c$ and 
the radius $R \cong 10^{-15} m$, which is the range 
of nuclear forces.

Expressing $N$ as $N = M/m$, $M$ total mass of the hadron, 
we obtain, following the usual
procedure, $A=M^{3/2}{\tilde A}$ and, finally, 
\begin{equation}
\alpha \cong (mc^{2})^{3/2}c^{-1}k^{-1/2}
R^{1/2} \, .
\end{equation}

\noindent Inserting numbers, we have again, up to at most
one or two orders of magnitude, $\alpha \cong h$.

\vspace{0.2cm}

This numerical equivalence with Planck constant
of the elementary unit of action 
per particle for any classical fundamental interaction on
each scale seems very significant, 
and can hardly be thought of being casual.
We again stress that one always obtains the same order of 
magnitude of $\alpha$ ($\cong h$) for any force law on its 
typical scale (universality of Planck constant).

Since the above procedure is of a grossly qualitative nature,
it seems important to provide other consistency checks of the 
fundamental Keplerian tremor law
eq. (5), to yield further support to its 
universal validity.

In fact, the natural question arises 
on how to determine the
order of magnitude, for each particular system,
of the characteristic time $\tau$. The latter must
obviously depend both on the universal 
elementary unit of action $\alpha \cong h$
and on the details of the chosen aggregate.

It is well known that the typical
time scale of quantum fluctuations 
is defined as the ratio between $h$  
and a suitable energy
describing the equilibrium state of the
given system on its characteristic dimensions. 
This leads naturally to identify this energy with
the thermal energy $k_{B}T$, with $k_{B}$ the
Boltzmann constant and $T$ the absolute temperature.

On the other hand, in our model such time scale
coincides with the fluctuative time $\tau$; we 
therefore write
\begin{equation}
\tau \cong \frac{h}{k_{B}T} \, .
\end{equation}

With the above definition we can rewrite the 
universal tremor hypothesis eq. (1) in the form:
\begin{equation}
T \cong \frac{h}{k_{B}}
\cdot \frac{\sqrt{N}}{{\cal{T}}} \, .
\end{equation}

We can adopt the point of view 
that the above equation 
be the definition of the 
absolute temperature for the system 
one is considering.
This definition, as one can see, 
connects the temperature to the global 
and the granular length scales,
and to the characteristic velocity 
associated to the given aggregate.

We now exploit this definition, 
applying it to some well 
established thermodynamic phenomena, 
as a further test of validity of 
our theoretical scheme.

\vspace{0.3cm}

{\it a) Emittance associated to charged beams in particle
accelerators}.
Among the stable aggregates of charged particles, a paradigmatic
role is played by the charged beams in particle accelerators.
Such systems are very interesting from our point of view
because they are generally described in classical terms, 
since they exist on a length scale that ranges approximately
from thirtytwo to thirtyfive orders of magnitude above the
Planck scale. 

However, our analysis shows that their stability, 
as in the gravitational case, is ruled by
Planck's constant, and it is thus ultimately 
of quantum rather than classical origin.

The bunch consists solely of charges of
the same sign, and stability (confinement) can be
achieved only through the action of an external 
focusing potential (magnetic field). 
Our analysis applies, for instance, by considering
the reference frame comoving with the
synchronous particle, yielding again $h$ as the unit of
action per particle 
(note that replacing electrons with protons does
not affect in a appreciable way the order of magnitude of
$\alpha$ due to the $m^{1/4}$ dependence in eq. (6)).

The emittance ${\cal{E}}$ is a scale of length
(or ,equivalently, of ``temperature'') associated to
charged beams, whose numerical value
in units of the Compton length $\lambda_{c}=h/mc$
lies, for typical accelerators (for instance
electron machines), in the range ${\cal{E}} 
\cong 10^{6} \lambda_{c} \div 10^{9} \lambda_{c}$
\cite{fedele}.
Following our scheme, we identify the characteristic
unit of emittance as the characteristic action
associated to the charged beam divided by $mc$.
The characteristic action associated to charged
beams is, in our framework,
$(k_{B} T) {\cal{T}} $, and therefore, by eq. (9),
the associated emittance is estimated as:
\begin{equation}
{\cal{E}} \cong
\lambda_{c} \sqrt{N} \, .
\end{equation}

We note that the above expression connects, at least
in the leading semiclassical order, the characteristic
emittance with the number of particles in a nontrivial
way.
\noindent Since in a typical bunch $N \cong 10^{11} \div
10^{12}$ \cite{particledata}, we finally 
obtain ${\cal{E}} \cong 10^{6} \lambda_{c}
\cong 10^{-6} m$
in good agreement with the phenomenological
order of magnitude estimated by other 
theoretical methods \cite{fedele}.

Therefore the Calogero model supplemented by eq.(8)
naturally supports a quantum--like
description of the dynamis of charged particle beams
with the correct order of magnitude for the emittance.

\vspace{0.2cm}

{\it b) Temperature of macroscopic systems}.
In this case
we know that $N \cong 10^{23} 
mol^{-1} \div 10^{24} mol^{-1}$
(Avogadro constant), 
and ${\cal{T}} \cong 10^{-2} s \div 10^{-3} s$,
corresponding to a rms 
velocity $v_{T} \cong 10^{2} m \cdot s^{-1}
\div 10^{3} m \cdot s^{-1}$ for 
gases around room temperature.
Inserting the numerical values of $h/k_{B}$
we obtain $T \cong 10^{2} K \div 10^{3} K$, 
as it should be.

\vspace{0.2cm}

{\it c) Temperature of quarks inside nucleons}.
In this case $N \cong 1$. The typical energy scale
$E$ of light quarks in a nucleon is of the order
of $\Lambda_{QCD} \cong 0.1 GeV$ \cite{particledata},
corresponding to a temperature $T = E/k_{B} \cong 
10^{12} K$. 
With the characteristic velocity of the order of
the velocity of light $c$, and the global 
scale of length $R \cong 10^{-15} m$, we have
${\cal{T}} \cong R/c \cong 10^{-23} s$.
Inserting numbers into 
eq. (9) we obtain just $T \cong 10^{12} K$.

\vspace{0.2cm}

{\it d) Bose--Einstein condensation}.
Recently, Bose--Einstein 
condensation has been experimentally
observed in a gas of 
rubidium and sodium atoms \cite{condensation}.
The condensate has linear dimension $R \cong
10^{-4} m$ at a temperature $T \cong 10^{-6} K$ 
and it contains $N \cong 10^{7}$ atoms. 
Letting ${\cal{T}} \cong R/v$,
with $v$ the characteristic velocity, eq. (9) yields  
\begin{equation}
v \cong \frac{k_{B}}{h} \cdot \frac{T}{\sqrt{N}} \cdot
R \cong 10^{-3} m \cdot s^{-1} \div 10^{-2} m
\cdot s^{-1} \, .
\end{equation}

The characteristic velocity thus is smaller by a factor
of the order $10^{-6}$ compared to the rms velocity of
the gas observed at room temperature, in agreement with
the theoretical prediction of a macroscopic ``zero''
momentum.

Therefore, the definition of temperature 
derived from the universal Keplerian tremor hypothesis
seems to be consistent.
We then move on to apply it to other
two significant cases.

\vspace{0.2cm}

{\it e) Cosmic background radiation}.
In the framework of our hypothesis it seems quite
reasonable to interpret the measured temperature 
associated to the cosmic background radiation,
$T = 2.7 K$, as the characteristic ``temperature
of the Universe''. Consequently, we insert in eq. (9)
a characteristic global time ${\cal{T}} \cong R/v $, 
with $R$ the Radius of the Universe and 
$v$ a characteristic velocity. This velocity cannot
be defined unambigously, therefore we take it in the
wide range that goes from $10^{5} m \cdot s^{-1}$ 
(the circular velocity of hydrogen clouds
surrounding galaxies) up to the velocity of light.

We now exploit eq. (9) to determine the order of
magnitude of $N$, the total number of particles in the
Universe. Inserting numbers:
\begin{equation}
N \cong 10^{66} \div 10^{72} \, ,
\end{equation}

\noindent which, in our crudely qualitative framework, is
compatible, within the error range,
with the value $N \cong 10^{\nu}$, $\nu = 78 \pm 8$,
estimated by cosmological arguments \cite{peebles}.

\vspace{0.2cm}

Some conclusive remarks are now due.
We first want to stress that,
according to our point of view, 
it is impossible to discriminate
the gravitational system, through the
observed Radius of the Universe $R$, from the other systems
(e.g. charged particles and quarks) by claiming that its
characteristic dimensions are not {\it a priori} 
determined by quantum mechanics while the 
dimensions of the other ones are.

In fact, one should note that there are about $20$
orders of magnitude separating the smallest scale of length
considered (that of the quarks confined in the nucleons)
and the Planck scale representing the fundamental quantum
mechanical scale of length.
Furthermore, if one considers {\bf macroscopic} aggregates 
like charged beams in particle 
accelerators, this difference
reaches up to $35$ orders of magnitude.

It seems then obvious to us that if one accepts, as it
is commonly asserted, that the influence of quantum
mechanics should manifest itself all the way through
this huge difference of length scales, it should
as well manifest itself also on the scale of length
of the Universe.

Therefore, the significant aspect that we single out in
the scheme put forward by Calogero is that it allows
for any dynamical system to obtain a quantum correction
to classical dynamics starting from a fluctuative
hypothesis of purely classical origin. 
This model seems then worth to be developed and
improved, since it could be of great conceptual
and computational relevance in the study of the
interplay between classical and quantum domains,
a fundamental issue in modern physics. 
We will report in a forthcoming paper how the
application of these ideas can be made already now
quantitative in the study of the quantum--like 
dynamics of charged particle beams in accelerators.

>From both the conceptual and the 
calculational point of view 
the next task would then be to
develop the present qualitative model into a fully
quantitative scheme of approximation 
(semi--quantal approximation scheme)
that would allow, in principle, to determine
all the higher--order quantum corrections, i.e.
a systematic reconstruction of quantum fetures 
from classical dynamics 
beyond the leading semiclassical order.
  
\vspace{0.4cm}

\noindent {\bf Aknowledgements}.

It is very hard for all of us to express in words 
our deep gratitude to Francesco Guerra for his
invaluable teachings and for his profound and 
far reaching insights, during our many--years
acquaintance, and in many enlightening 
discussions on all aspects of the present 
work. 
We also gratefully aknowledge very
useful conversations with Francesco
Calogero on his model and on an early draft of 
the present paper.

One of us (F.I.) aknowledges 
the Alexander von Humboldt--Stiftung for 
financial support and the Fak\"ult\"at f\"ur
Physik der Universit\"at Konstanz for
hospitality while on leave of absence from
the Dipartimento di Fisica dell'Universit\`a
di Salerno.

\end{document}